\documentclass[prb,aps]{revtex4}
\usepackage{amssymb}
\usepackage{amsmath}
\usepackage{colordvi}
\usepackage[latin1]{inputenc}
\usepackage[brazil,portuges]{babel}
\usepackage{graphicx}
\begin{document}

\title{A semiquantitative approach to the impurity-band-related
transport properties of GaMnAs nanolayers}

\author{E. J. R. de Oliveira}
\affiliation{Instituto de Física, Universidade do Estado do Rio de Janeiro,
Rio de Janeiro, RJ, Brazil}
\author{E. Dias Cabral}
\affiliation{Universidade Estadual da Zona Oeste, Rio de Janeiro, RJ, Brazil}
\author{M. A. Boselli}
\affiliation{Instituto de Física, Universidade Federal de Uberlândia,
Uberlândia, MG, Brazil}
\author{I. C. da Cunha Lima}
\thanks{Author to whom correspondence should be addressed}
\email{ivandacunhalima@pq.cnpq.br}
\affiliation{Instituto de Física, Universidade do Estado do Rio de Janeiro, Rio de Janeiro, RJ, Brazil}

\date{\today}
\begin{abstract}
  We investigate the spin-polarized transport of GaMnAs nanolayers in which a
ferromagnetic order exists below a certain transition temperature. Our calculation
for the self-averaged resistivity takes into account the existence of an impurity
band determining the extended (``metallic'' transport) or localized (hopping by thermal
excitation) nature of the states at and near the Fermi level. Magnetic
order and resistivity are inter-related due to the influence of the spin polarization of
the impurity band and the effect of the Zeeman splitting on the mobility edge.
We obtain, for a given range of Mn concentration and carrier density, a ``metallic''
behavior in which the transport by extended carriers dominates at low temperature,
and is dominated by the thermally excited localized carriers near and above the
transition temperature. This gives rise to a conspicuous hump of the resistivity which has been
experimentally observed and brings light onto the relationship between transport and
magnetic properties of this material.
\end{abstract}
\pacs{75.50.Pp,72.25.Dc,72.15.Rn,72.20.Ee}

\maketitle
\section{Introduction}

Diluted magnetic semiconductors (DMS) are important materials in Spintronics since, in their ferromagnetic phase, they can be used as a source or a filter of  spin-polarized carriers  to be injected into semiconductor heterostructures. It is interesting to have the transition temperature as high as possible, aiming operation at room temperature. In the study of the possible origins for the magnetic order, the existence of an impurity band has been questioned. An impurity band occur when impurities, donors or acceptors, are close enough to have a finite width of the density of states (DOS) for the bound electron or hole. It is well known that at high concentration the maximum of DOS approaches the conduction band (donor impurities) or the valence band (acceptor impurities), and at the same time a tail appears in the extended state band (conductance or valence). At impurity concentrations still higher the two DOS coalesce, and the energy gap between then disappears. The Mott-Anderson transition occurs when the states at the Fermi level change their character from localized to extended. The Fermi level is said, then, to have crossed an edge of mobility, separating localized and extended states. Although, at least up to present, this transition cannot be classified as a phase transition with known critical exponents etc, the concept of the mobility edge is well accepted. If it plays an important role in the magnetic properties of DMS, it is a question of identifying somehow the existence of an impurity band itself.

Ga$_{1-x}$Mn$_x$As is a semiconductor alloy belonging to the DMS family. Example of DMS materials are among oxides (like ZnO doped with Eu, Co etc), II-VI compounds (like CdTe doped with Mn), III-V compounds (like GaMnAs), and IV-VI compounds (like PbTe doped with Eu). The possibility of controlling the energy gap of the non-magnetic alloys GaAlAs and InGaAs, the material's $n$ or $p$ character by doping, the electron-hole recombination time, the effective mass, among other properties, together with the capability of growing these materials in heterostructures of high quality, makes GaMnAs special. After the work of van Esch \textit{et al},\cite{vanEsch} and Oiwa  \textit{et al},\cite{Oiwa} where they  produced good homogeneous samples by low-temperature molecular beam epitaxy and obtained ferromagnetic phases with high transition temperatures ($T_C$), many improvements brought this temperature to 180K in epilayers.\cite{olejnik} Lower $T_C$ were obtained in multilayer system, which are important for devices.\cite{trilayer} Recently, much higher $T_C$ near 250 K were obtained in highly doped digital layers.\cite{tanaka,tanaka1}  Although higher transition temperatures have been obtained in other DMS, the characteristics given above for GaAs focus the attention on GaMnAs. Besides of its importance for Spintronics applications, it shows challenging questions concerning the understanding of its magnetic and transport properties.

Taking as a basis for discussion those samples in Refs. \onlinecite{vanEsch} and \onlinecite{Oiwa},
a non-metal-to-metal transition is observed as the Mn concentration increases and goes above $x=0.03$. This is
followed by a metal-to-non-metal transition near $x=0.065$. On the other hand, the ferromagnetic
phase is observed even before the first transition, but in this case its ferromagnetic-to-paramagnetic
transition temperature is low. At first this temperature increases with the Mn concentration, reaching a maximum inside
the concentration range corresponding to the metallic phase. For concentrations above the
one giving rise to the metal-to-non-metal transition, ferromagnetism persists but with decreasing
ferromagnetic-to-paramagnetic transition temperature. Recently, the  quality of the samples
increased still further, leading to the possibility of doping GaAs with higher concentrations of Mn and obtaining higher Curie temperatures. The conductive behavior, concerning the non-metal-to-metal transition and
vice versa, together with its influence over the transition temperature are still observed in these samples.\cite{olejnik,chiba08} In Mn doped InAs quantum wells resistivity measurements have also shown localization effects in the hole system.\cite{wurtbauer10}

During the last ten years a lot of attention was given to understand the mechanism
leading to the ferromagnetic order in GaMnAs. Several possibilities were explored, such
as RKKY interaction,\cite{bose0} mean field approximation,\cite{dietl} magneto-polaron \cite{bahtwolf}
etc. We have performed \cite{bos1} self-consistent
Monte Carlo simulations using an RKKY-like model to obtain the magnetization,
susceptibility and Curie temperature of several GaMnAs/GaAs heterostructures. We
concluded that the possibility of creating an inhomogeneous spin-polarization density across the
sample, more specifically, having a highly concentrated polarization density inside the magnetic layer,
increases remarkably the Curie temperature. More recently, Green's function
formalisms in different approaches were used to treat this problem. One of them uses a Matsubara
Green's function and modeled the spin correlation by a Brillouin function to obtain the
conductive behavior on ferromagnetic quantum dots.\cite{lebedeva_prb} Also,
the Kondo lattice model was used to treat the diluted regime in the ferromagnetic
phase in the DMS,\cite{nolting} as well as a local random phase approximation.\cite{bouzerar}
All these methods consists of treating disorder in the magnetic interaction, but neglect Coulomb scattering by impurities. As a consequence, localization is not considered. On the other hand, a survey on the optical and transport properties of GaMnAs samples over a wide range
of concentrations \cite{nottingham,prague} identified the existence of an impurity
band and explored the character of the states near the Fermi level.\cite{manyauthors}
This latter, based on photoluminescence  measurements, found that even in samples of Ga$_{1-x}$Mn$_x$As doped with very low Mn concentration, i.e., in non-metallic samples, there are evidences of the formation of an impurity band. For metallic samples with a doping $> 2\%$, no effect of an isolated impurity band was perceived in terms of dc transport, suggesting a fusion of the impurity band with the valence band. Impurity bands in GaMnAs have been also suggested after weak localization observation.\cite{furdyna} Calculations of impurity band have also been performed.\cite{usapl,fiete} It is known that the character of holes determines the indirect interaction \cite{bhat1,nolting}
among the magnetic ions, influencing the Curie temperature of the magnetically ordered
samples. Therefore, our motivation in this paper is to explore two aspects that we
consider as key issues to understand the inter-related magnetic and transport
properties in GaMnAs: spin polarization and localization.

For decades the nature of the states of the non-interacting and the interacting quasi-two-dimensional electron systems
involving, for instance, the metal-to-non-metal transition, the possibility of having extended
states in absence of an external magnetic field, and the existence of a minimum metallic conductivity
crossed periods where things seemed to be settled, alternating with periods where new questions
and observations appeared. For the sake of having a historical view of those questions, one can
follow the discussions appearing in Refs. \onlinecite{Andloc} and \onlinecite{Krav}. In the present case, carriers
are holes, their concentration is much higher than in typical  Mott-Anderson transition in semiconductors
(in the order of 10$^{18-19}$cm$^{-3}$, while in Ga$_{0.95}$Mn$_{0.05}$As it is around
10$^{20}$cm$^{-3}$). Besides, the spin-polarization in GaMnAs is due to the local potential interaction, generally accepted to be modeled by an sp-d Kondo-like potential, instead of being due to an external magnetic field. Disorder enters in the random Coulomb scattering by impurities, and also on the sp-d term, in the latter due to the stochastic
interaction with the fluctuating local dipole moments together with the random location of these
dipoles.

Recently the effects of the disorder in a
quasi-two-dimensional GaMnAs layer was treated using a self-consistent multiple-scattering
approximation to calculate its density of states and spectral density function. Using
parameters corresponding to GaMnAs thin layers, a wide range of Mn concentrations and hole
densities were explored to understand the nature, localized or extended, of the spin-polarized
holes at the Fermi level for several values of the average magnetization of the Mn system.
For a certain
interval of Mn and hole densities, an increase on the  magnetic order of the Mn ions comes
together with a change of the nature of the states at the Fermi level, showing a
delocalization of spin-polarized extended states anti-aligned to the average Mn magnetization,
and a higher spin-polarization of the hole gas. This was associated to the appearance of the
non-metal-to-metal transition caused by the increase of Mn concentration, once the ferromagnetic
phase is reached, and was observed experimentally since the first samples were produced. The
metal-to-non-metal transition observed in samples of Refs. \onlinecite{vanEsch} and
\onlinecite{Oiwa} appears in that calculation as Anderson transitions occurring in the spin
anti-aligned impurity band of samples with higher Mn concentration.

In this paper we link the magnetic and the transport properties coming out of a picture in
which the existence of an impurity band of spin-polarized holes determines the localized or
extended nature of states at and near the Fermi level. At a given average magnetization,
for a specific sample, the Zeeman-like splitting, as described below, determines a mobility edge.
Then, two transport mechanism take place.
One, a thermally activated hopping of the localized states, those occupied states lying below the mobility edge in the impurity band. Another, the ``metallic'' transport, corresponding to the contribution of the occupied extended states, whose energies
lie between the mobility edge and the Fermi level. We have, therefore, two channels like parallel circuits, and
if we describe the transport by the resistance, it will be given by law of the inverses. The
calculation herein proceeds as follows. Section II shows the method used to obtain the density
of states, how we proceed to the classification of the states as extended or localized, and in consequence, how we obtain a mobility edge. Section III describes a method allowing directly the calculation of the self-averaged
``metallic'' resistance. Then, Section  IV applies these formalisms to a double layer sample providing a highly
inhomogeneous spin polarization density, seeking for an ideal combination of Mn concentration
and density of holes. Finally Section V compares our results with experimental observations and
makes general comments based on these results.

\section{Quasi-two-dimensional impurity band and the character of the hole states}
\label{figure-of-merit}

Our system is described by the following Hamiltonian:
\begin{equation}
H= \sum_{{\bf p}\sigma}\epsilon_p a^{\dag}_{{\bf p}\sigma}a_{{\bf p}\sigma}
+\sum_{{\bf q}\sigma}U_{\textrm{imp}}({\bf q})\sum_{{\bf p}} a^{\dag}_{{\bf p}+{\bf
q}\sigma}a_{{\bf p}\sigma}+ \sum_{{\bf q}\sigma\sigma'}V_{\textrm{mag}}({\bf q},\sigma,\sigma')\sum_{{\bf p}} a^{\dag}_{{\bf p}+{\bf
q}\sigma}a_{{\bf p}\sigma'}.
\label{secquat}
\end{equation}
The second  term on the right represents the Coulomb scattering by impurities, and the third term is the interaction of a hole  with the localized magnetic moments at
the Mn positions. It is modeled by a Kondo-like potential:
\begin{equation}
V_{sp-d}(\vec r)=-I\sum_i\vec S_i\cdot \vec s(\vec r)\delta(\vec
r- \vec R_i), \label{hund}
\end{equation}
where the localized spin of the Mn ion $\vec S_i$ at position
$\vec R_i$ is treated as a classical variable, and $\vec s(\vec
r) $ is the spin operator of the carrier at position $\vec r$; $I$
is the $sp-d$ interaction. The third term on the right represents the Coulomb scattering by impurities, eventually the same ions giving rise to the local dipole moments.

The impurity band is obtained by averaging the Green's function (GF) on the dipole moment orientation and on the impurity position,  which, for a given scatterer configuration is defined by:
\begin{equation}
G^R({\bf p},t;{\bf p}',t')= -i\left<0|T[a_{{\bf
p}}(t)a^{\dag}_{{\bf p'}}(t')]|0\right>.
\label{greenfunc}
\end{equation}
Taking a diagrammatic expansion we reach a Dyson equation for the averaged GF by including all
irreducible insertions to form the self-energy $\Sigma^R(\bf
p,E)$. The configurational averaged GF (which is
diagonal in $\bf p$, as a consequence of the space becoming
homogeneous after the averaging process) can be written in terms
of the Dyson equation:
\begin{equation}
<G_{\sigma}^{R}({\bf p};E)>=G_0^{R}({\bf p};E)
+G_0^{R}({\bf p};E)\Sigma_{\sigma}^R({\bf
p},E)<G_{\sigma}^{R}({\bf p};E)>.
\end{equation}
Obtaining this average is an extremely complicate task, which has not been achieved yet. Our choice has been to treat separately the magnetic and Coulomb scattering. The first one is taken in zero-th order, neglecting fluctuations of the
magnetic moments on the impurities, which otherwise produce
spin-flip scattering. Besides, we neglect fluctuation on the magnetization and assume
homogeneous distribution of the localized magnetic moments with a density given by the Mn
concentration with an average magnetization $<S>=5\hbar/2<M>$, where $ \langle M \rangle $ is the normalized magnetization
($0\leq \langle M \rangle  \leq 1$) . This leads to a real contribution to the self-energy which is ${\bf p}$-independent, and can be incorporated to the energy particle
\begin{equation}
\epsilon_{\sigma}({\bf p})=\epsilon_{0}({\bf p})-\frac{x}{2}N_0\beta  \langle M \rangle \sigma,\label{ansatz0}
\end{equation}
 This approximation corresponds to the zero-th order term in Ref. \onlinecite{lebedeva_prb}. $N_0\beta x=Ic$, $c$ stands for the Mn density, $\sigma=\pm 1$, $x$ is the Mn doping
factor, and $N_0\beta$ is the exchange potential for holes,
$N_0\beta=-1.2eV$, according to Ref. \onlinecite{matsukura}.

\begin{figure}[h]
 \includegraphics[width=10cm,angle=-90]{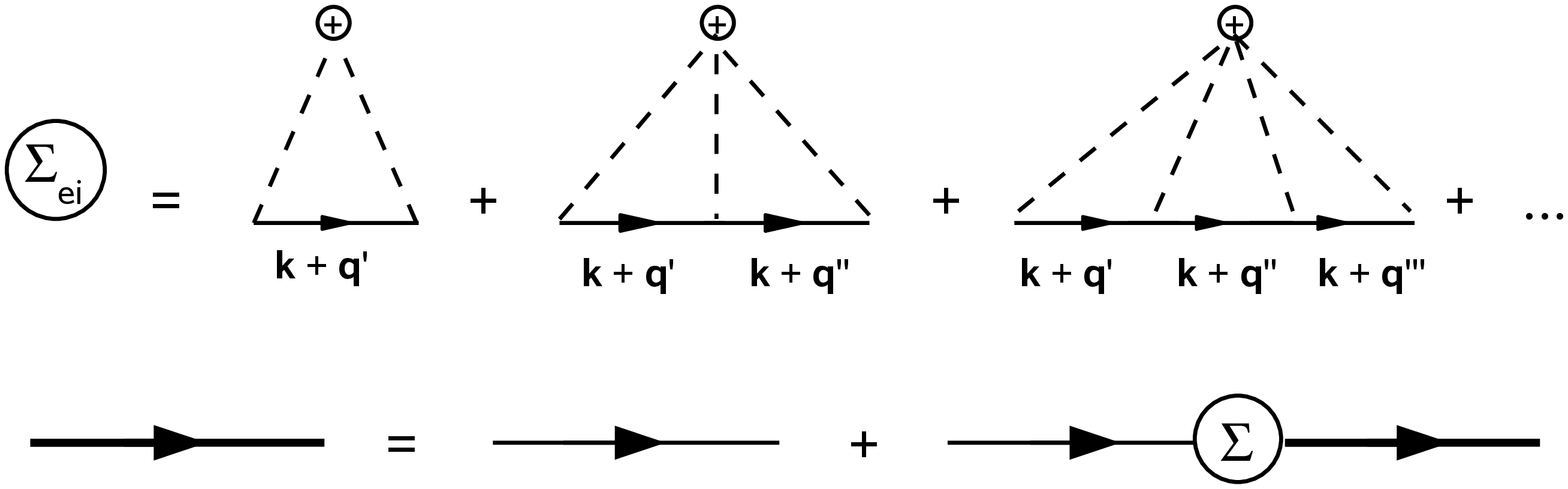}
\caption{Self-energy diagrams, within the multiple scattering
approximation, and the Dyson equation for the averaged GF.} \label{selfea}
\end{figure}

Next we proceed to the calculation of the self-energy due to impurity scattering. In order to simplify the notation we will omit the subscript $\sigma$, since the derivation is performed for each spin orientation individually. Due to the high impurity density, the calculation of the average Green's function must consider terms beyond the self-consistent Born approximation. This problem has been handled successfully in the past with the use of the multiple-scattering approximation which
consists in considering self-energy insertions due to repeated scattering by the same impurity. Here we use the method developed by   Klauder \cite{klauder} with the improvements performed by Ghazali and Serre.\cite{ghaz1}
The multiple scattering approximation consists in selecting from
the self-energies only those terms  described by the diagrammatic
expansion in Fig. \ref{selfea}, expressed by the expansion:

\begin{eqnarray}
\Sigma_{ei}({\bf p};E)=&&\frac{1}{(2\pi)^d}\int d^d{\bf q_1}v({\bf
q}_1)G^0({\bf p}+{\bf q}_1)v(-{\bf q}_1)\nonumber \\
&&+\frac{1}{(2\pi )^{2d}}\int d^d{\bf q}_1 \int d^d{\bf q}_2 v({\bf
q}_1)G^0({\bf p}
+{\bf q}_1)v({\bf q}_2-{\bf q}_1)G^0({\bf p} + {\bf
q}_2)v(-{\bf q}_2)+ ...
\end{eqnarray}
Next, we define the vertex function $K({\bf k},{\bf q}_1;E)$, which obeys the following Dyson equation:

\begin{equation}
K({\bf k},{\bf q}_1;E)=\frac{1}{(2\pi)^d}\int d^d {\bf q}'_1
v({\bf q}_1-{\bf q}'_1)\overline{G({\bf q}'_1)}
[N v({\bf k}-{\bf q}'_1)+K({\bf k},{\bf q}'_1;E)].
\label{voila}
\end{equation}
It can be shown, as derived in Refs.\cite{ghaz1,jsupercondnovmagn_nosso}, that the self-energy can be obtained from the vertex function by:
\begin{equation}
\Sigma_{ei}({\bf k};E)=K({\bf k},{\bf k};E).
\label{sigk1}
\end{equation}
So, $\Sigma_{ei}({\bf k};E)$ is the
diagonal term corresponding to line ${\bf k}$, row ${\bf k}$ of
matrix  $K$, calculated at energy $E$. For the sake of making the formalism
operational, a function
$U$  is defined such that:
\begin{equation}
U({\bf k},{\bf q};E)=K({\bf k},{\bf q};E)+N v({\bf k}-{\bf q}).
\label{defU}
\end{equation}
In that case, Eq.(\ref{voila}) becomes:
\begin{equation}
U({\bf k},{\bf q};E)=N v({\bf k}-{\bf q})
+\frac{1}{(2\pi)^2}\int d^d {\bf q}'_1 v({\bf q}'-{\bf
q})\overline{G({\bf q}')} U({\bf k},{\bf q}';E).
\end{equation}
With this result we have a linearized matrix equation:

\begin{equation}
[I-\widetilde{v}\widetilde{G}]\widetilde{U}=N\widetilde{v},
\label{linearized}
\end{equation}
where the sign tilde is used to identify a matrix. Therefore, the
problem  is reduced to a simple linearized matrix equation of the
kind $\tilde{A} \cdot \tilde{X}=\tilde{B}$, where
$\tilde{B}=N\tilde{v}$ is a fixed matrix, while
$\tilde{A}=[I-\tilde{G}\tilde{v}]$, and $\tilde{X}=\tilde{U}$
changes at each iteration in the self-consistent process.

If $x=0.05$,
$V_{mag}$ introduces a kind of Zeeman splitting of 150 meV for
fully magnetic ordered samples. It is important to emphasize the
fact that the splitting in two bands, one with spins aligned with
the average magnetization (which goes up in energy) and the other
with spins anti-aligned (which goes down in energy), is not really
a Zeeman splitting since it does not result of the interaction of
the spin with an external magnetic field, but it comes out of  a
configurational average performed on the sp-d interaction of the
Kondo-like local potential.

The self-consistent calculation is performed for a two-
dimensional hole gas of areal density $n_s$ submitted to Coulomb
scattering by a negative ionized impurity system of concentration
$N_i$. These two
data, $n_s$ and $N_i$, are considered as independent parameters.
This is important in the present context, since it is known that
the density of free carriers in the ferromagnetic GaMnAs samples
is just a fraction of the concentration of Mn, the reason for this
being, still, a controversial question in the literature. Besides, we leave the method open to the possibility of doping with non-magnetic dopants, or changing the carrier concentration by applying a gate voltage.
We start the self-consistent calculation with a free-particle spin-polarized Green's
function $G_{\sigma}^0(\textbf{k},E)$:
\begin{equation}
G_{\sigma}^0(\textbf{k},E)=\frac{1}{E-\epsilon_{\sigma}({\bf k})+ i0^+}.
\end{equation}
Next, we calculate the density of states as a function of the
impurity  and carrier concentrations. A deeper
understanding on the localized/extended character of the states can be attained by analyzing the
spectral density function (SDF),
\begin{equation}
A_{\sigma}(k,E)=-\frac{1}{\pi} Im[<G_{\sigma}^{R}({\bf k};E)>]
\end{equation}
at a given energy as a function of the wave vector. Then, we
observe the typical shapes corresponding to localized and extended
states, as described in the following discussion.

An electron bound to an impurity
in the very diluted regime has a bell-shaped SDF with the maximum
at $k=0$. A free particle, having its energy perfectly determined by the
wave vector, $E(k)=\hbar^2k^2/2m$ has its SDF at a given $E$ represented by
delta function in variable $k$ centered at $k=\sqrt{2m^*E}/\hbar$.
When random scatterers are included, a width is introduced in the density of states of the single impurity bound state, together with a tail in the bottom (top) of the conduction (valence) band. As the impurity concentration increases, the isolated impurity band merge into the tail, and the gap between impurity states and free carriers disappear.
The  shape of the SDF for a given energy is used as a
semi-quantitative criteria for the identification of a state at energy $E$ as being
extended or localized. For instance, if the energy is deep inside
the conduction/valence band the SDF plotted as a function of the
wave vector $k$ shows a very sharp distribution at a given $k$,
while a state deeply inside an impurity band appears as a broad
Lorentzian centered at $k=0$. This is schematically shown in
Fig. \ref{sdfscheme}. At left, (a) shows the DOS of a characteristic
two-dimensional n-doped semiconductor system with $n_s=2\times10^{12}$ cm$^{-2}$ and $N_i=5\times
10^{11}$ cm$^{-2}$ (intermediate concentration
regime), with a well defined hole impurity band and the valence
band with its very small tail, not visible at this concentration. At right, (b) shows the SDF obtained for the
two values of energy indicated on (a). The energy $E_1$ lies in
the middle of the impurity band, while $E_2$ lies deep inside the
conduction band. This is demonstrative of the fact that knowing the SDF at the Fermi level we can
infer the character of our sample model as metallic or non-metallic according to the shape of the SDF.

For each spin polarization the mobility edge is identifyed to the energy in which the SDF changes the shape from a sharp Lorentzian to a bell-shaped curve at $k=0$. We have obtained this transition to be quite evident, the change occurring inside a narrow range of energy, as compared to the value of the Fermi level. Once the
spin-dependent mobility edge is found for a given $<M>$, we obtain the density of extended
carriers for each metallic sample:
\begin{equation}
n^*_s=\sum_{\sigma}\int_{\mu_{\sigma}}^{E_F} dE {\cal
N}_{\sigma}(E).\label{freecharge}
\end{equation}
The difference $n_s-n^*_s$ represents the density of localized holes in the sample. We have, then, two conduction channels established: one, a ``metallic'' channel corresponding to the transport of the extended holes with energy lying between the mobility edge and the Fermi level, with density $n_s^*$; another, that obtained by thermal activation of the localized states with activation energy $\Delta=|E_F-\mu|$. The total resistance is given by
\begin{equation}
\frac{1}{R_{\textrm{total}}} =\frac{1}{R_{\textrm{hopping}}}+\frac{1}{R_{\textrm{metallic}}}.
\end{equation}
The first term on the right hand represents the resistance due to a hopping process, modeled by
\begin{equation}
R_{\textrm{hopping}}=R_{\infty}\exp(\Delta/T).\label{hopping}
\end{equation}
The method we used to calculate the ``metallic'' resistivity is described in the next Section.

\section{Resistivity of the ``metallic'' channel}
We assume  quasi-two-dimensional carriers of density $n^*_s$ (arial density) inside a layered semiconductor structure were impurities (scatterer centers) are located inside specific layers with a density $n_i$.  The calculation is performed according to the Lei and Ting's \cite{leiting0} force and energy balance equation formalism which has been used successfully in several heterostructures of different dimensionality \cite{lei1,lei2,ivanwang1} both in the linear and in the non-linear regime. An advantage of this formalism is to provide a direct calculation of the self-averaged resistivity, instead of the conductivity.

In the present case, a confining potential $U(z)$ due to the structure interfaces breaks the symmetry in the z-direction, and in the absence of impurities and of an electric field, the single-particle states in the envelope function approximation are solutions of the Schrödinger equation:
\begin{equation}
H_{0}\Psi_{n{\bf{k}}}({{\bf{r}}},z)={\cal E}_{n{\bf{k}}}\Psi_{n{\bf{k}}}({{\bf{r}}},z),\label{envfunc}
\end{equation}
resulting into a free motion in the $(x,y)$ plane and the envelope function  $\phi_{n}(z)$ in the $z$-direction:
\begin{equation}
\Psi_{n{\bf{k}}}({{\bf{r}}},z)=\frac{1}{\sqrt{A}}e^{i{\bf{k}}\cdot{{\bf{r}}}}\phi_{n}(z),\label{autof}
\end{equation}
with sub-band energies:
\begin{equation}
{\cal E}_{n{\bf{k}}}={\cal E}_{n}+\frac{\hbar^{2}k^2}{2m^{*}}.\label{autoe}
\end{equation}
Equation (\ref{envfunc}) is solved in the self-consistent Hartree-Fock approximation.

Including scattering by impurities and the external electric field, the total Hamiltonian in the $(x,y)$ plane can be separated into center of mass (CM) and relative coordinates of the $N$ carriers,
\begin{equation}
H=H_{cm}+H_e+H_{ei},\label{htotal}
\end{equation}
where
\begin{eqnarray}
&&H_{cm}=\frac{P^2}{2Nm^{*}}-Ne{{\bf{R}}}\cdot\bf{E}\label{h2qcm}\\
&&H_e=\sum_{m,\bf{k}\sigma}{\cal E}_{n{\bf{k}}}c^{\dagger}_{m\bf{k}\sigma}c_{m\bf{k}\sigma}\\
&&H_{ei}=\sum_{{m,n,\sigma}\atop{{\bf{k}},{\bf{q}},a}}U_{mn}({\bf{q}},z_a)e^{-i{\bf{q}}\cdot({\bf{R}}+{\bf{r}}_{a})}c^{\dagger}_{m{\bf{k}}+{\bf{q}}\sigma}c_{n{\bf{k}}\sigma}\label{h2qbi}
\end{eqnarray}
Vectors $\mathbf{P}$ and  $\mathbf{R}$ represent the momentum and in-plane position of the CM. We have used the creation and destruction operators  $c^\dag_{m{\bf{k}}\sigma}$($c_{m{\bf{k}}\sigma}$),
and $U_{mn}({\bf{q}},z_a)$ represents the potential matrix element of the interaction of an ion at position $(\mathbf{r}_a,z_a)$. It is  given by:
\begin{equation}
U_{mn}({\bf{q}},z_a)=\frac{Ze^2}{2\pi\epsilon_{0}\kappa{q}}\int{dz}\phi_{m\sigma}^{*}(z)\phi_{n\sigma}(z)e^{-q|z-z_a|},\label{potbi}
\end{equation}
where $\kappa$ is the GaAs dielectric constant.

The configurational averaged drag force due to  impurities depends on the polarization function:
\begin{equation}
\Pi_{mn\sigma,m'n'\sigma'}({\bf{k}},{\bf{k}}',{\bf{q}};t-t')=
\mathrm{T}\langle[c^{\dagger}_{m{\bf{k}}'-{\bf{q}}\sigma}(t)c_{n{\bf{k}}'\sigma}(t),c^{\dagger}_{m'{\bf{k}}+{\bf{q}}\sigma'}(t')c_{n'{\bf{k}}\sigma'}(t')]\rangle_{0}.\label{eqpi}
\end{equation}
where $\mathrm{T}$ is the time ordering operator. Defining $\omega_{d}\equiv{\bf{q}}\cdot{\bf{v}}_{d}$, the averaged drag force is written as:
\begin{equation}
{\bf{f}}_{i}(v_d)=\frac{-in_{i}}{\hbar}\sum_{mn{\bf{q}}}\sum_{m'n'}{{\bf{q}}}\int{dz_a}U_{mn}({\bf{q}},z_a)U_{m'n'}(-{\bf{q}},z_a)h(z_a)\nonumber\\
\sum_{{\bf{k}}\sigma}\sum_{{\bf{k}}'\sigma'}\Pi_{mn\sigma,m'n'\sigma'}({\bf{k}},{\bf{k}}',{\bf{q}};\omega_{d}),\label{fitot2}
\end{equation}
with $f(z_a)$ equals one if $z_a$ lies inside the impurity layer, zero otherwise, and
\begin{equation}
\Pi_{mn\sigma,m'n'\sigma'}({\bf{k}},{\bf{k}}',{\bf{q}};\omega_{d})=
\int_{-\infty}^{+\infty}dt'e^{i\omega_{d}(t-t')}\Pi_{mn\sigma,m'n'\sigma'}({\bf{k}},{\bf{k}}',{\bf{q}};t-t')\label{pitrf}\nonumber \end{equation}
We have neglected hole-hole correlation, therefore the polarization function is diagonal. In this case,
\begin{equation}
{\bf{f}}_{i}(v_d)=\frac{n_{i}}{\hbar}\sum_{mn{\bf{q}}}\bigl|U_{mn}(q)\bigr|^{2}{\bf{q}}{\Pi}_2^{(0)}(m,n,{\bf{q}},\omega_{d}),\label{fimpw}
\end{equation}
with the effective potential due to impurities
\begin{equation}
\bigl|U_{mn}(q)\bigr|^{2}=\int{dz_a}\bigl|U_{mn}({\bf{q}},z_a)\bigr|^{2}h(z_a)\label{pefimp}
\end{equation}
and
\begin{equation}
{\Pi}_2^{(0)}(m,n,{\bf{q}},\omega_{d})\equiv\Im \sum_{{\bf{k}}\sigma}\frac{f_{0}(\xi_{m{\bf{k}}+{\bf{q}}})-f_{0}(\xi_{n{\bf{k}}})}{\xi_{m{\bf{k}}+{\bf{q}}}-\xi_{n{\bf{k}}}-\omega_{d}-i\delta}\label{defpi}
\end{equation}
is the imaginary part of the non-interacting polarization function.
In the stationary regime the balance of force on the system gives
\begin{equation}
n_{s}^{*}e{\bf{E}+{\bf{f}}}_i(v_d)=0.\label{ebfee}
\end{equation}
In that case, we have for the resistivity
\begin{equation}
\rho_i=-\frac{{\bf{f}}_i\cdot{\bf{v}}_d}{(n_{s}^{*}ev_d)^2}.\label{resimp}
\end{equation}

Assuming ${\bf{v}}_{d}=v_{d}\hat{x}$ and isotropy, the resistivity becomes,
using $\omega_{d}=v_{d}q_{x}$
\begin{equation}
\rho_i=-\frac{n_{i}}{2\hbar{n_{s}^{*}}^{2}e^2}\sum_{mn{\bf{q}}}\frac{q^2}{\omega_{d}}\bigl|U_{mn}(q)\bigr|^{2}{\Pi}^{(0)}_{2}(m,n,{\bf{q}},\omega_{d}).\label{rho2}
\end{equation}
In the Ohmic regime we take $\omega_{d}\rightarrow{0}$, and
\begin{equation}
\rho_i=-\frac{n_{i}}{2\hbar{n_{s}^{*}}^{2}e^2}\sum_{mn{\bf{q}}}\frac{q^2}{\omega_{d}}\bigl|U_{mn}(q)\bigr|^{2}
[{\Pi}^{(0)}_{2}(m,n,{\bf{q}},0)+\dot{\Pi}^{(0)}_{2}(m,n,{\bf{q}},0)\omega_{d}],\label{rhoohm}
\end{equation}
where
\begin{equation}
\dot{\Pi}^{(0)}_{2}(m,n,{\bf{q}},0)\equiv[\frac{\partial{\Pi}^{(0)}_{2}(m,n,{\bf{q}},\omega_{d})}{\partial\omega_{d}}]\Biggr|_{\omega_{d}=0}.\\
\end{equation}

\section{Impurity band and transport}

\begin{figure}[h]
\includegraphics[width=8cm,angle=-90]{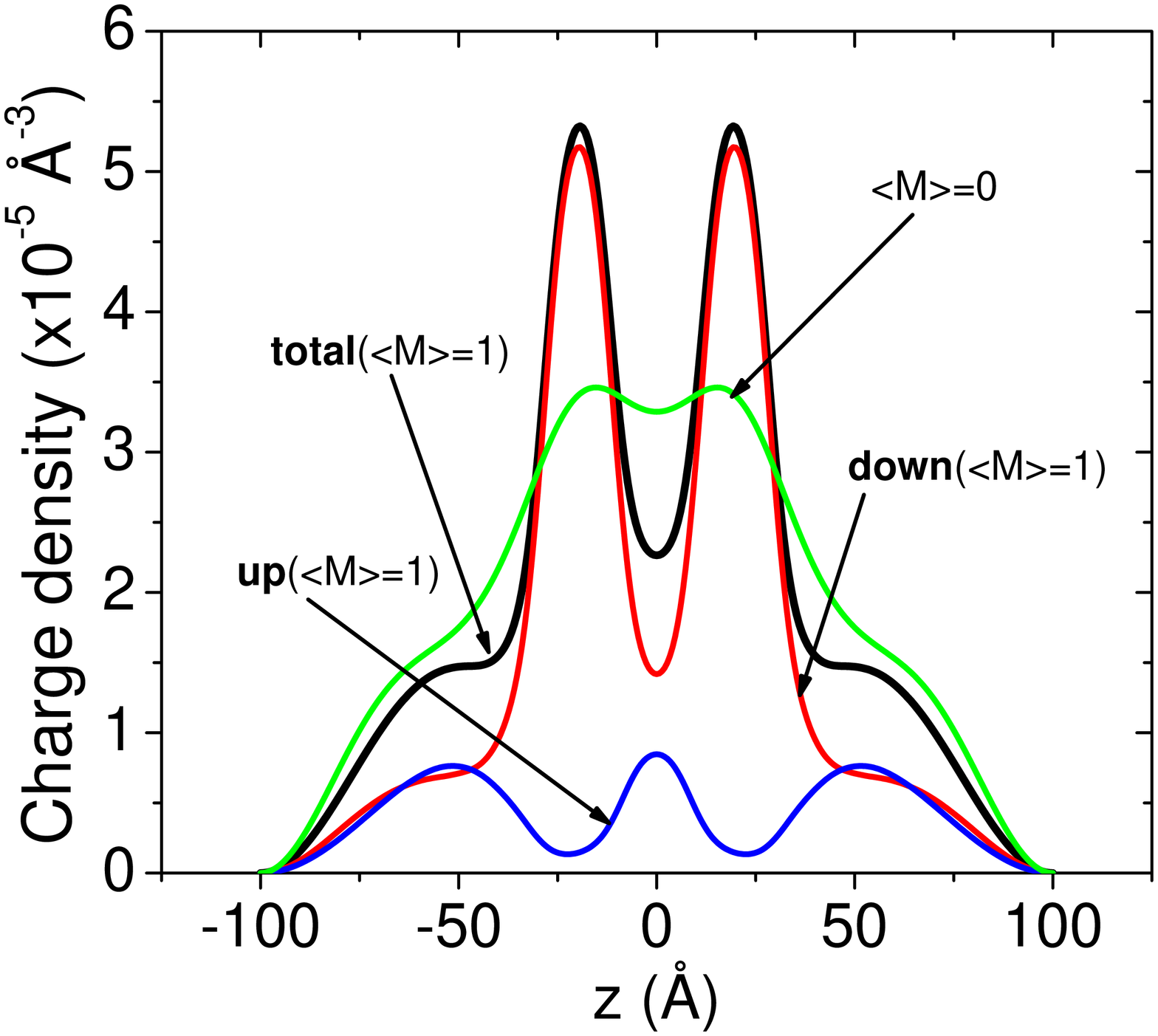}
\caption{(color online) Charge distribution across the z-direction for $\langle M \rangle=0$ and  $\langle M \rangle=1$.} \label{chargedensity}
\end{figure}

In what follows we present the calculation for the resistance of a double layer system,
composed by two GaMnAs layers of width 20\AA\  immersed in a wide GaAs host. These two layers are separated by a GaAs spacer of width 20\AA . First we assume that we have only the ``metallic''  channel, and we fix the carrier density while changing the magnetization. In that case, for a given carrier concentration $n_s$, we observe a redistribution of the charge density in the z-direction as the average magnetization $<M>$ decreases from $<M>=1$ to $<M>=0$, shown in the two extreme cases by Fig. \ref{chargedensity}, where $n_i=1.\times10^{20}$ cm$^{-3}$ and $n_s=2\times10^{13}$ cm$^{-2}$.
The resistivity is expected to decrease as the average magnetization decreases, a consequence of the carriers occupying a wider region than that occupied by the impurities. In other words, we should expect the scattering cross-section to decrease towards the paramagnetic phase.

\begin{figure}[h]
\includegraphics[width=8cm,angle=-90]{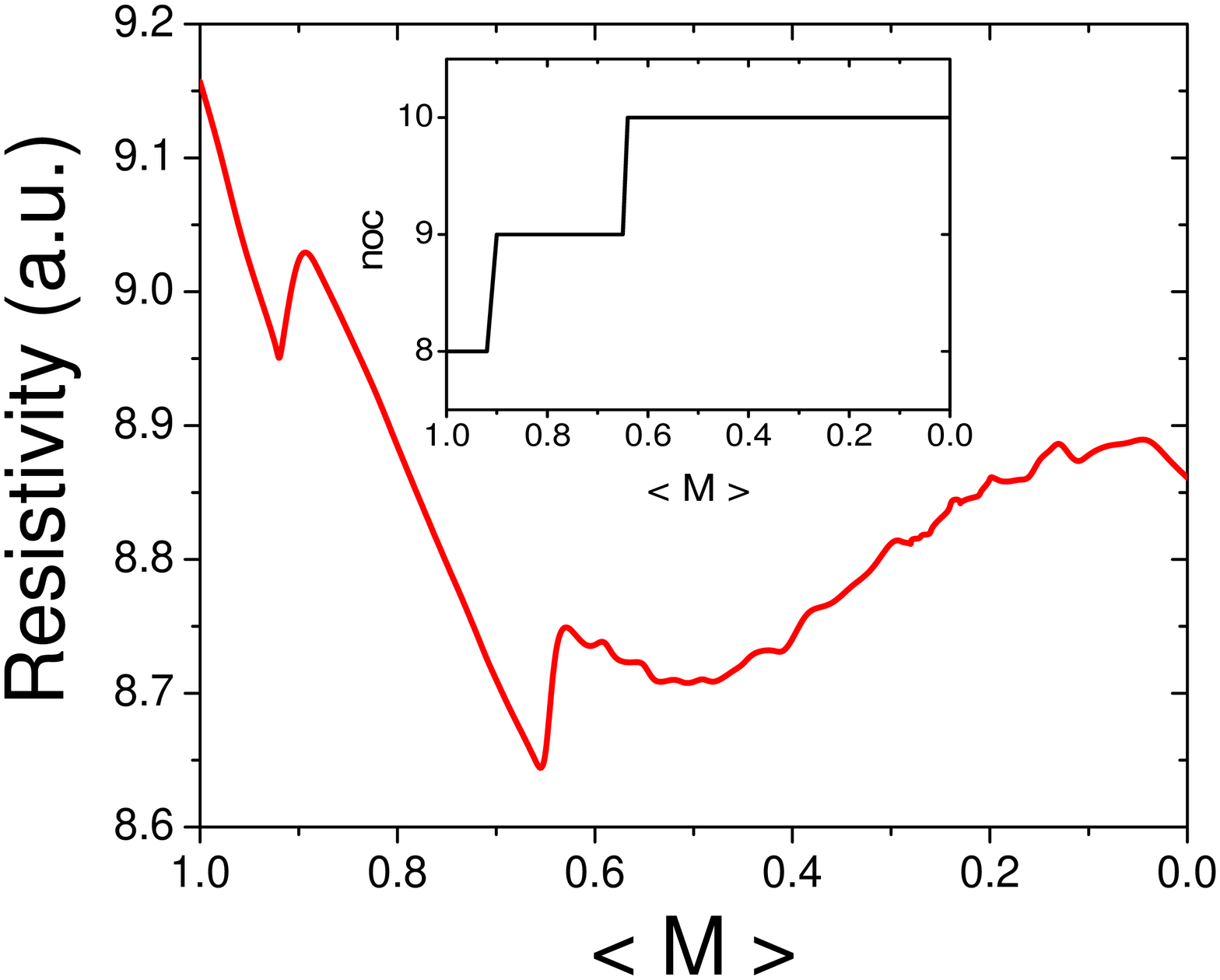}
\caption{(color online) Resistivity as a function of the average magnetization for fixed carrier density. The scale is chosen in such a way that the temperature corresponding to a given magnetization grows to the right. The inset shows the index of the sub-band containing the Fermi level.} \label{naive}
\end{figure}

At the same time as the sub-band occupation changes with $\langle M \rangle$, the Fermi level moves, as shown in the inset of Fig. \ref{naive} jumping suddenly from a weakly occupied sub-band to a strongly occupied one. A similar behavior is observed as an effect of changing an applied magnetic field, so we should obtain the equivalent to the Shubnikov-de Haas oscillations. This is exactly what is observed in Fig. \ref{naive} showing the resistivity (notice the linear scale) as a function of the average magnetization. However, as $\langle M \rangle \rightarrow 0$ several sub-bands become occupied, the spin-polarization disappears, several sub-bands of both spin orientation tends to contribute to the polarization function in equal foot. This provides an increase of the resistivity near $\langle M \rangle=0$. In conclusion, if we maintain the carriers density constant as we decrease the average magnetization, we obtain in our calculation oscillations of the resistivity in the range of strong magnetic order, followed by a minimum of the resistivity, and a  hump as we approach $\langle M\rangle =0$.

\begin{figure}[h]
\includegraphics[width=8cm,angle=-90]{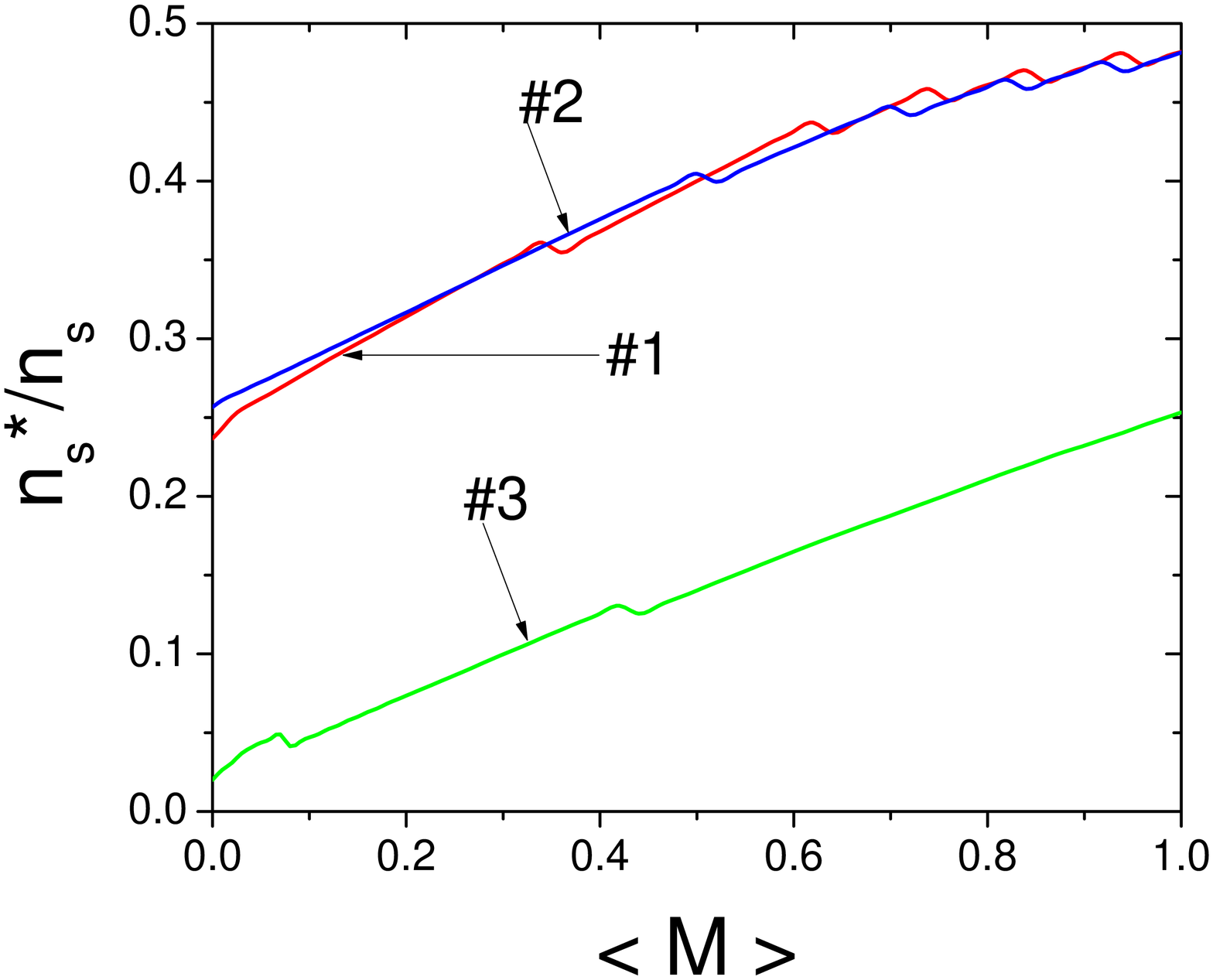}
\caption{(color online) Dependence of the ratio $n_s^*/n_s$ with $<M>$ for three different samples; sample $\#1$: $n_s=3.\times 10^{19}$cm$^{-3}$ and $n_i=3.75\times 10^{19}$cm$^{-3}$; sample $\#2$: $n_s=3.25\times 10^{19}$cm$^{-3}$ and $n_i=4.5\times 10^{19}$cm$^{-3}$; sample $\#3$: $n_s=3.25\times 10^{19}$cm$^{-3}$ and $n_i=7.5\times 10^{19}$cm$^{-3}$.} \label{nsns*}
\end{figure}

Next we consider the existence of an impurity band, appearing as a wide region where a non-zero density of states occurs below the edge of the valence band, superimposed to a tail of the valence band density of states. The signature of this impurity band shall be expressed by the existence of two transport channels, one due to eventually localized states, other to eventually extended states.
The total density of particles, in our case holes, is assumed fixed. So, as the magnetization changes we have $n_s^*(\langle M\rangle)$ for the arial density of the extended states, and $n_s-n_s^*$ for the localized states.
The ratio $n_s^*/n_s$ is obtained by the following ansatz.
For a given pair of parameters $(n_i,n_s)$ labeling the sample, the spectral density function is used to determine the effective mobility edge according to the criteria described in Section \ref{figure-of-merit}. Knowing
the mobility edge
we calculate directly the $n_s^*/n_s$ ratio for a given $n_i$, as a function of the magnetization, as shown in Fig. \ref{nsns*} for three different samples.
In sample $\#1$ we have $n_s=3.\times 10^{19}$cm$^{-3}$ and $n_i=3.75\times 10^{19}$cm$^{-3}$, i.e, $n_s$ and $n_i$ are chosen very close to each other. Sample $\#2$ has $n_s=3.25\times 10^{19}$cm$^{-3}$ and $n_i=4.5\times 10^{19}$cm$^{-3}$. Finally sample $\#3$ has $n_s=3.25\times 10^{19}$cm$^{-3}$ and $n_i=7.5\times 10^{19}$cm$^{-3}$, $n_s$ as a small fraction of $n_i$.

Notice that in sample $\#3$ the ratio goes to approximately zero when the magnetization goes to zero, describing a sample with a very high resistance in the ``metallic'' channel (almost as an insulator) near $T_C$. The other two samples also show a decrease of the ratio as $<M>\rightarrow 0$, but remaining ``metallic'' near $T_C$.

Now, given a sample $n_i$ we assume the ratio $n_s^*/n_s$ to follow the same dependence as obtained in the 2D limit, and we take also the same activation energy, $\Delta= |E_F-\mu|$, which is a reasonable approximation if we are not looking for quantitative comparison. On the other hand, $R_{\infty}$ in Eq. (\ref{hopping}) is calculated by fitting the resistivity of samples $\#1$, $\#2$ and $\#5$ in Ref. \onlinecite{Oiwa} to the function:
\begin{equation}
R_{\infty}=\frac{\alpha}{x^{\beta/3}},
\end{equation}
for a given Mn concentration $x$. We expected the dependence to be nearly proportional to the inverse of the average distance between Mn atoms. In fact we obtained $\beta=1.4$. Then we interpolated the hopping resistance for the necessary Mn concentration.

\begin{figure}[h]
\includegraphics[width=8cm,angle=-90]{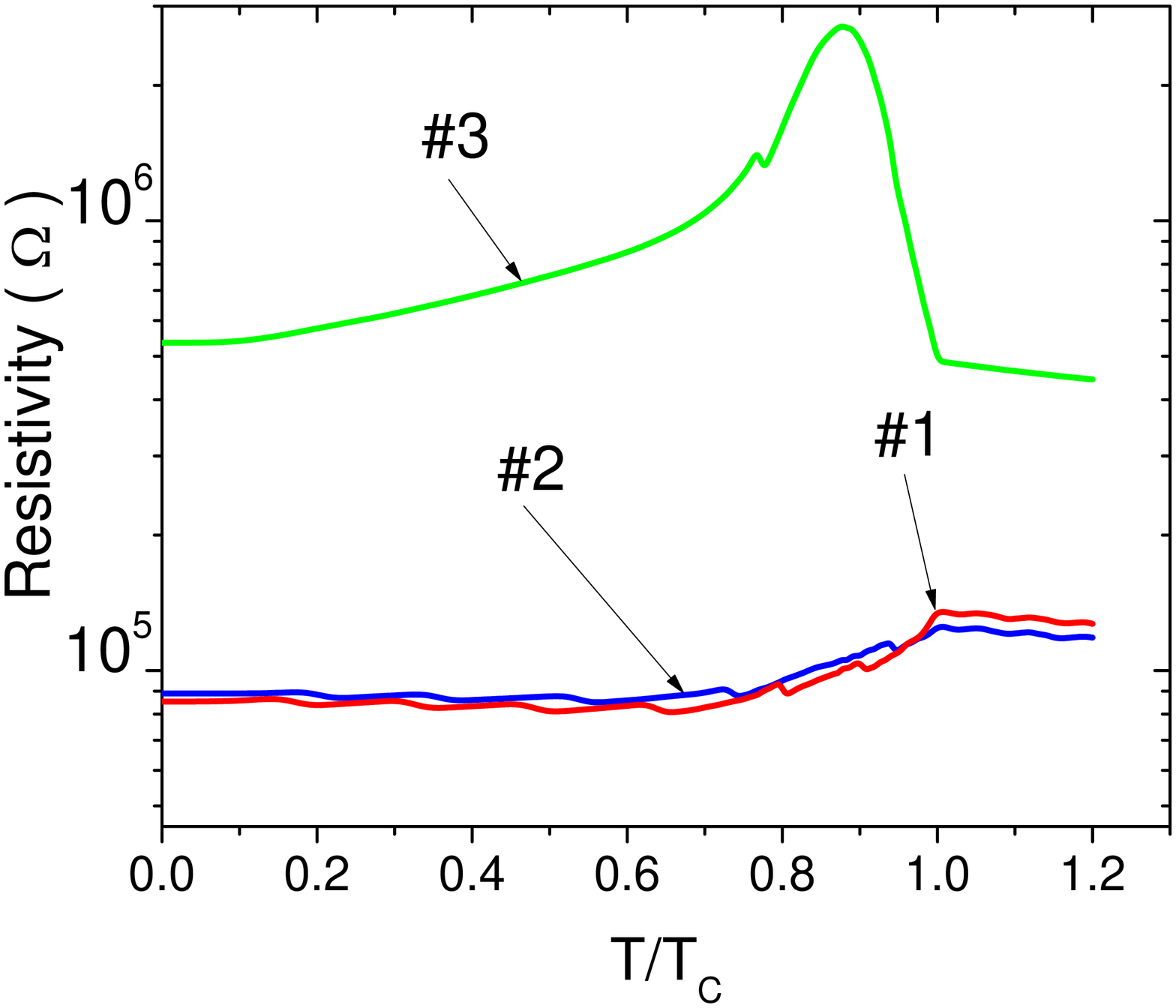}
\caption{(color online) Total resistivity as a function of temperature for the three samples described above.} \label{totalresist}
\end{figure}

The calculation of the total resistivity for the same three samples is shown in Fig. \ref{totalresist}, in logarithmic scale. Notice that curves for samples $\#1$ and $\#2$ are similar, showing a maximum of the resistivity  at $T_C$, and a lower  value of the resistivity across the whole temperature range, as compared to sample $\#3$, the one which behaves like an  insulator near $T_C$. This latter shows a pronounced hump just before $T_C$ and a flat region at low temperatures. This sample contains the clear signature of the impurity band. In fact, at low temperature, on the flat region, it shows a typical  metallic behavior. Notice that the dependence of $n_s^*$ on $<M>$, together with the effect of finite temperature, fades the  Shubnikov-de Hass oscillations. As we approach $T_C$ the chemical potential gets close to the mobility edge, raising in  importance the transport via thermal excitation of the localized states.

\begin{figure}[h]
\includegraphics[width=10cm,angle=-90]{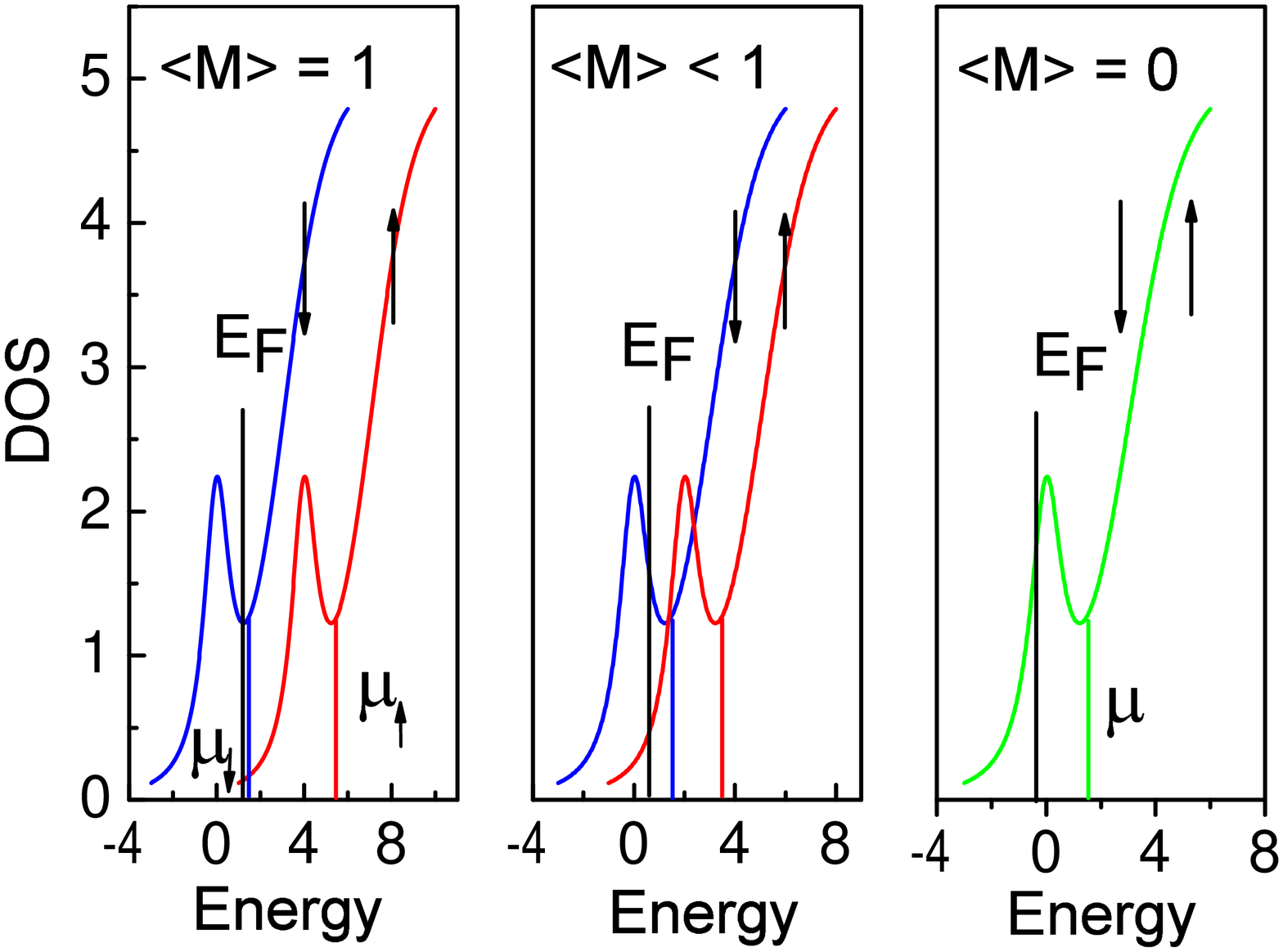}
\caption{(color online) Schematic classification of samples as metallic or nonmetallic.
The downward (upward) arrow indicates the anti-aligned (aligned) spin impurity band and its corresponding mobility edge. The Fermi levels lies, in this case, to the left of the anti-aligned mobility edge and the sample is non-metallic in the whole range of average magnetization.}
\label{esqimpA}
\end{figure}

\begin{figure}[h]
\includegraphics[width=10cm,angle=-90]{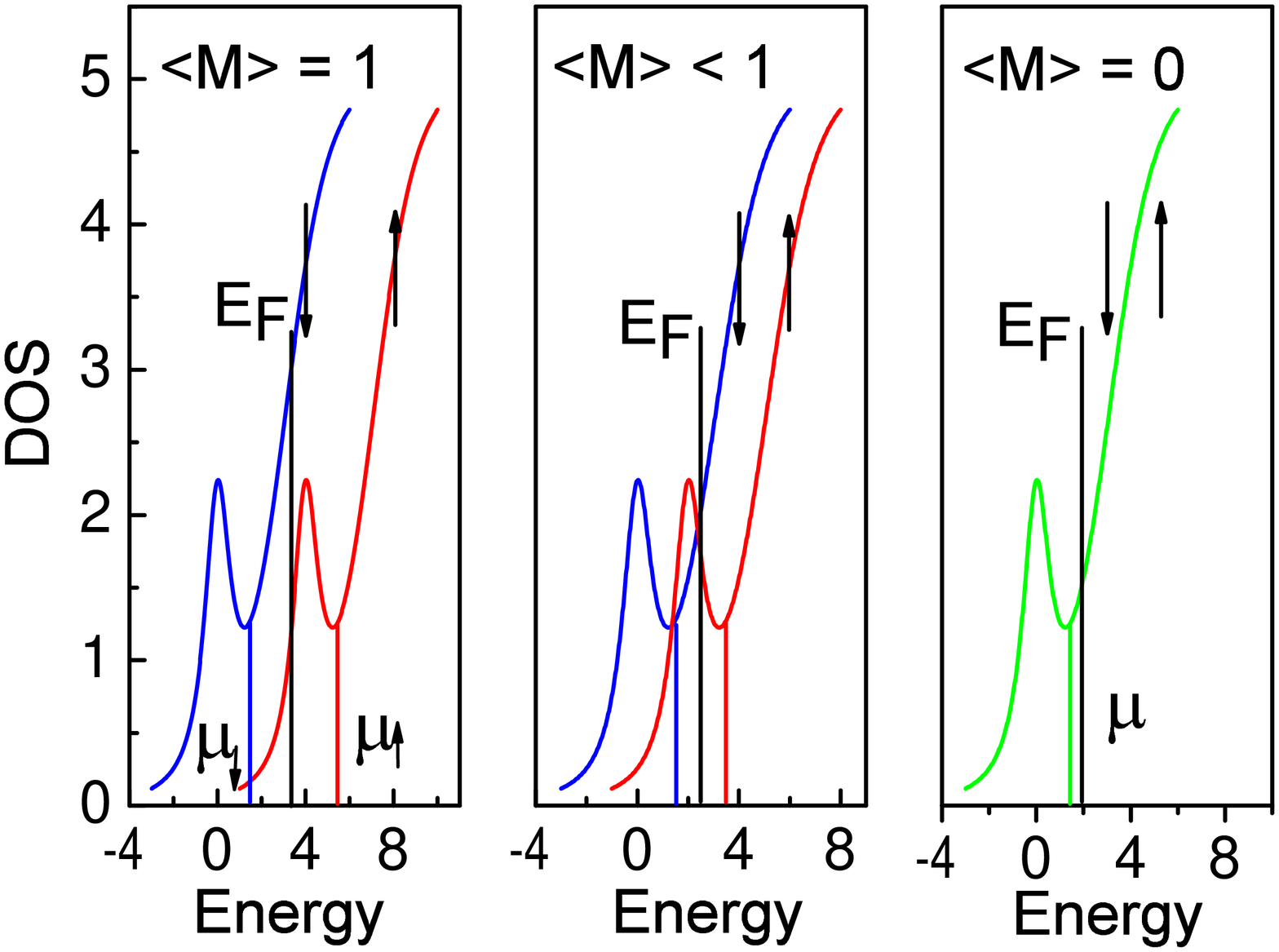}
\caption{(color online) As before, but the Fermi level is kept to the right of the anti-aligned mobility edge, even as $<M>$ goes to zero. This is the case of a sample which is metallic for the whole range of the magnetization.}
\label{esqimpB}
\end{figure}

\begin{figure}[h]
\includegraphics[width=10cm,angle=-90]{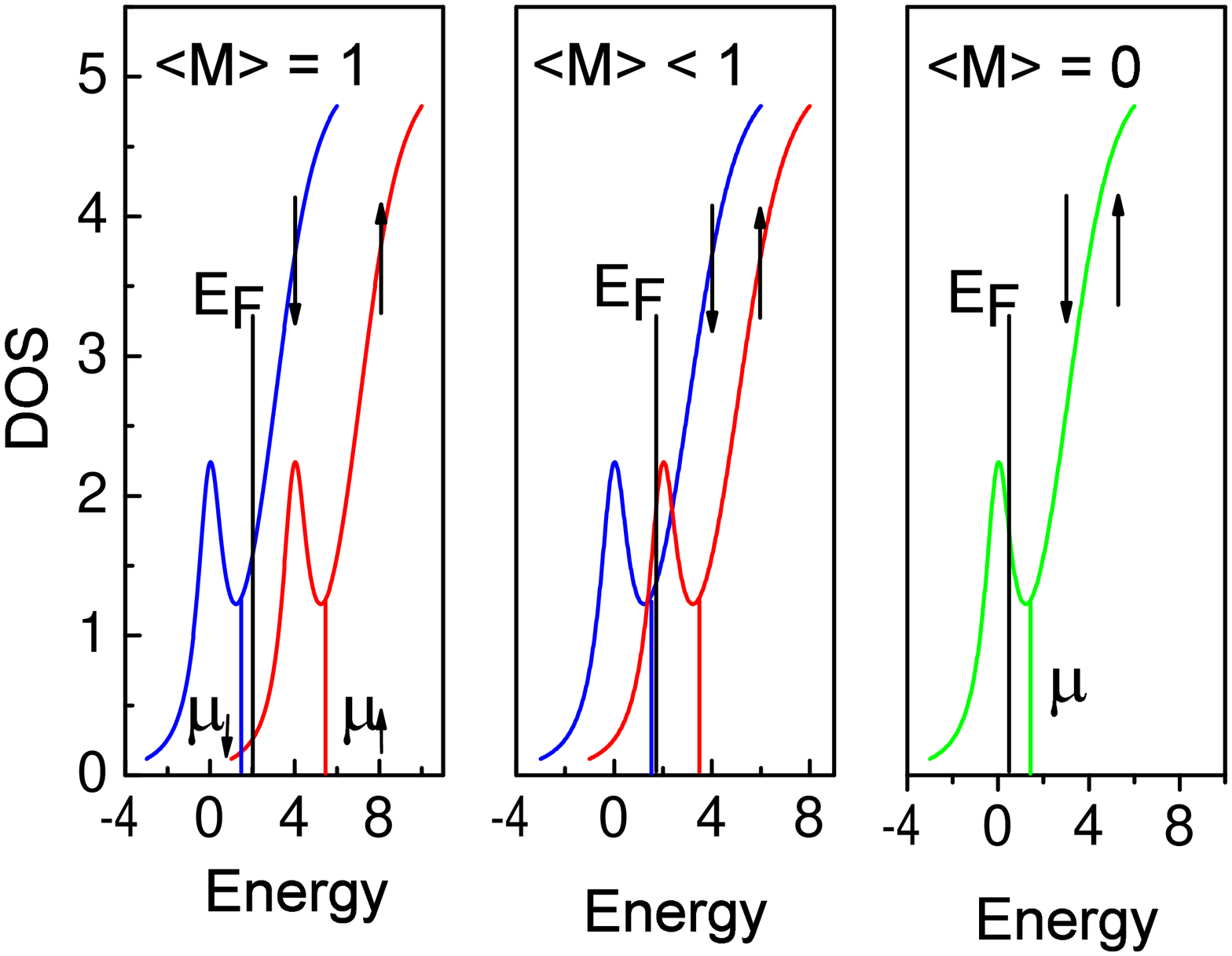}
\caption{(color online) As before, but in this case the mobility edge lies to the right of the anti-aligned mobility edge at $<M>=1$, but crosses to its left as $M$ goes to zero.}
\label{esqimpC}
\end{figure}

To end our discussion about the character of metallic or non-metallic Ga$_{1-x}$Mn$_x$As samples , Figs. \ref{esqimpA}-\ref{esqimpC} show schematically the relative positions of the impurity and valence band for three values of magnetization and three different situations as regard the classification of samples as metallic or nonmetallic. In this figure, the solid line curves represent, respectively, the DOS of  anti-aligned and aligned spins with the magnetization. The vertical lines  represent their mobility edges  and the Fermi level, which moves to the left as the average magnetization goes to zero, once  we fixed the left edge of the anti-aligned DOS.

In Fig. \ref{esqimpA}  we take a sample in which the Fermi level lies to the left of the anti-aligned mobility edge. This samples is an insulator even at $<M>=1$. As $<M>$ decreases the splitting decreases and the Fermi level goes to the left, keeping all the states localized. In Fig. \ref{esqimpB} the carrier density is high enough so that even with the Fermi level going to the left as the magnetization decreases, it is maintained to the right of the anti-aligned mobility edge. This sample behaves as a metal even at $<M>=0$, like samples  $\#$ 1  and  $\#$ 2. Fig. \ref{esqimpC} represents a sample which is ``metallic'' at $<M>=1$, but as the magnetization decreases the Fermi level moves to the left of the spin anti-aligned mobility edge, and the sample behaves as non-metallic below a certain $<M>$.

It is remarkable that the schematic case of Fig. \ref{esqimpC} is typically the situation observed by the calculation of the resistivity of sample $\#$ 3 in Fig. \ref{totalresist}, containing the signature of the impurity band. Notice that, since we are treating a quasi-two-dimensional sample, the density of states just after the tail of the valence band has a very sharp derivative, as can be seen in the real calculation of Fig. \ref{sdfscheme}. This means that the Fermi level is expected to be located just after this tail, if the sample has a metallic behavior. In that case, depending on the carrier concentration, when the Fermi level moves to the left as the magnetization decreases, it may occupy the region below the valence band tail, in other words, a position already inside the impurity band. In that case, a change of sign on the derivative of the DOS should be observed. It is important to mention that this is in line with the explanation for the observation of anomalous Hall effects performed in GaMnSb in Ref. \onlinecite{lou}.

 \section{Conclusion}

\begin{figure}[h]
\includegraphics[angle=-90,width=10cm]{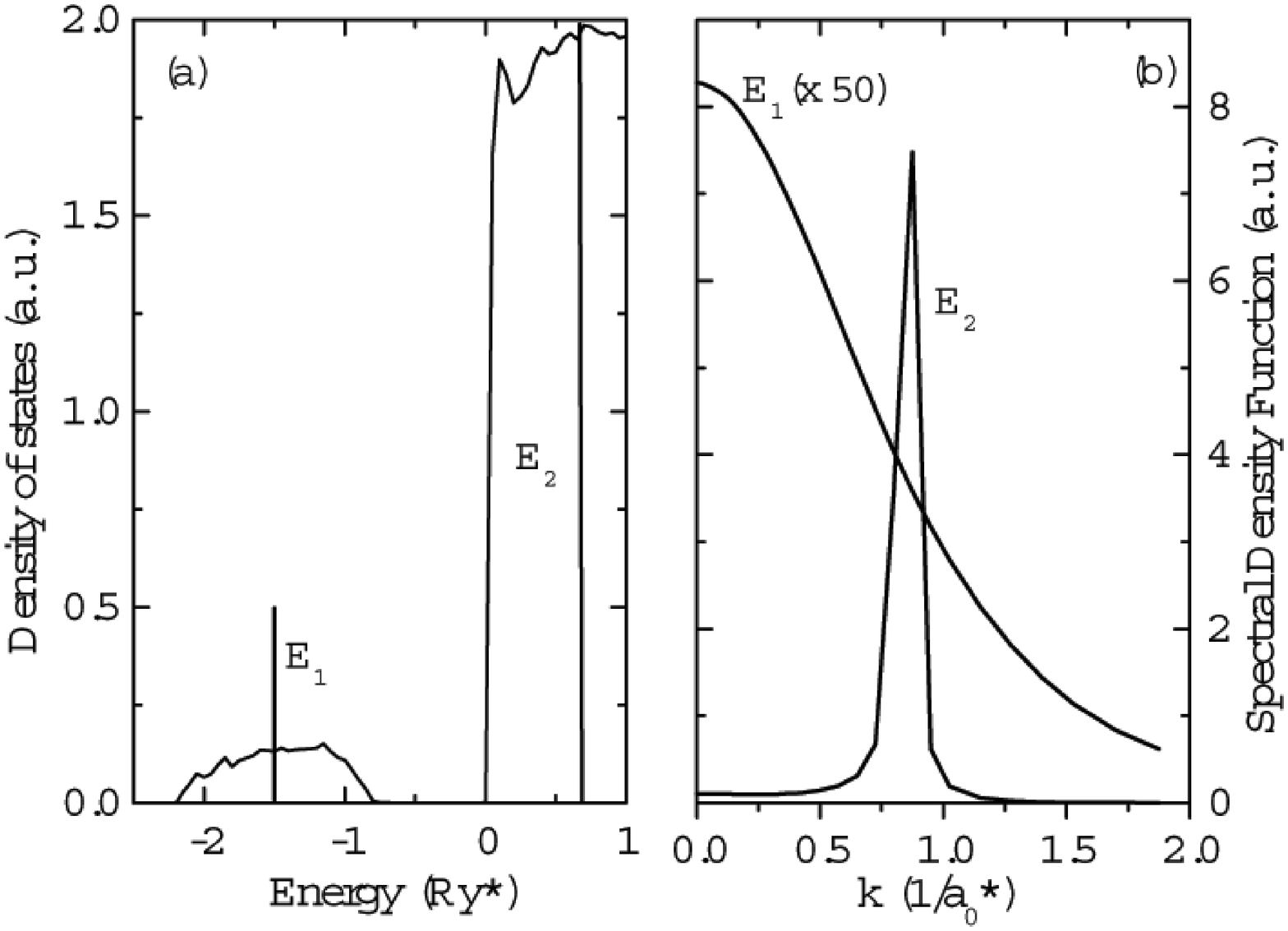}
\caption{Typical spectral density function for localized states (like that with energy $E_1$) and extended states (like that with energy $E_2$).} \label{sdfscheme}
\end{figure}

 These calculations present theoretical evidence of the existence of an impurity band contribution to the transport when the curve of the resistivity versus temperature has a typical metallic behavior at low temperature and a pronounced hump with a maximum just before the transition temperature. Although the calculations have been performed in a double layer structure, where the effects of spin-polarization are enhanced, and, in consequence, the occurrence of metallic to a non-metallic behavior as the critical temperature is approached, we believe that our conclusion can be extended from the digital up to the epilayer case, and is an additional evidence, together with optical measurements, of the importance of a hole impurity band in GaMnAs. It is worthwhile to add that the existence of an impurity band creates not only two transport channel, but two interaction mechanism between localized magnetic moments as well: one via localized states (possibly a magneto-polaron mechanism), another via extended states (RKKY, for instance). Samples like the one represented schematically by Fig. \ref{sdfscheme} would have both mechanism acting up to a certain value of the temperature, above which the low average magnetization makes the mobility edge to lie above the Fermi level, and the ``extended'' states channel ceases acting. From that point on, the magnetization decreases faster with temperature. We believe this is the reason for which it is hard to approximate the curve magnetization as a function of temperature in layered GaMnAs structures by a simple Brillouin function. \cite{trilayer}

\newpage

\end{document}